\documentclass[aps,prb,twocolumn,amsmath,showpacs]{revtex4}%

\usepackage{bm}
\usepackage{graphicx}

\begin{document} 
\title{Alternating current loss 
	in radially arranged superconducting strips}
\author{Yasunori Mawatari$^{\ \rm a)}$}
\affiliation{%
	Energy Technology Research Institute, 
	National Institute of Advanced Industrial Science and Technology, \\
	\hspace*{6em}
	Tsukuba, Ibaraki 305--8568, Japan
	\hspace{6em} 
}
\author{Kazuhiro Kajikawa}
\affiliation{%
	Research Institute of Superconductor Science and Systems, 
	Kyushu University, \\
	6--10--1 Hakozaki, Higashi-ku, Fukuoka 812--8581, Japan
}
\date{February 16, 2006}

\begin{abstract}
Analytic expressions for alternating current (ac) loss in radially arranged 
superconducting strips are presented. 
We adopt the weight-function approach to obtain the field distributions 
in the critical state model, and we have developed an analytic method 
to calculate hysteretic ac loss in superconducting strips for 
small-current amplitude. 
We present the dependence of the ac loss in radial strips upon the 
configuration of the strips and upon the number of strips. 
The results show that behavior of the ac loss of radial strips 
carrying bidirectional currents differs significantly from that 
carrying unidirectional currents. 
\end{abstract}
\maketitle

Alternating current (ac) losses are of great importance for electric power 
applications of superconducting wires. 
Most of the commonly used high-temperature superconducting wires 
have tape or strip geometry, and theoretical expressions for hysteretic 
ac losses for a single superconducting strip have been derived by 
Norris~\cite{Norris70} for ac transport currents and by 
Halse~\cite{Halse70} for ac magnetic fields, 
based on the critical state model.~\cite{Bean62} 
See also Ref.~\onlinecite{Brandt93}. 
Although electromagnetic interaction among multiple strips must be considered 
in multifilamentary wires and electric power devices, only a few studies 
have been carried out to derive analytic expressions for ac loss in 
multiple strips, e.g., in arrays containing an infinite number of 
strips~\cite{Mawatari96} or in two coplanar strips.~\cite{Ainbinder03} 

\begin{figure}[b] 
\includegraphics{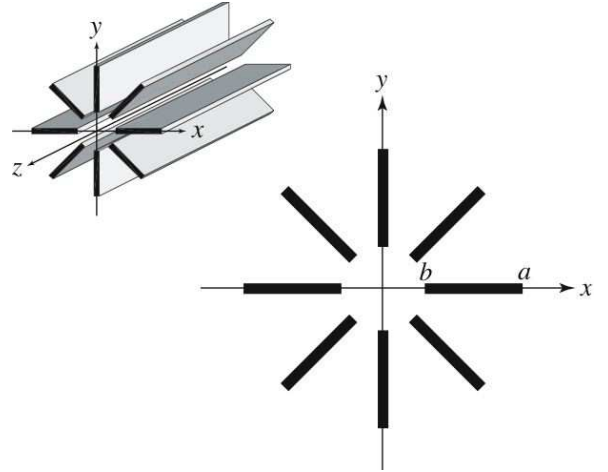}
\caption{%
Configuration of radially arranged superconducting strips. 
The $k$th strip is at $b<\rho<a$ and $\theta=\vartheta_k\equiv 2\pi k/n$, 
and carries a transport current $I_k$ [where $\rho= (x^2+y^2)^{1/2}$, 
$\theta= \arctan(y/x)$, and $k=1,2,\cdots,n$]. 
Positive integer $n$ denotes the number of strips, and this figure 
shows an example for $n=8$.
}\label{Fig_radial-strips}
\end{figure}

In this letter, analytic expressions are derived for ac loss 
in another type of multiple strips, namely, radially arranged strips. 
In the radial strips, each superconducting strip has width $a-b$ 
and thickness $d$ (where $a>b\gg d$ and $a-b\gg d$), 
and is infinitely long along the $z$ axis. 
The critical current of each strip is given by $I_c=j_cd(a-b)$, 
where $j_c$ is the constant critical current density.~\cite{Bean62} 
The strips of number $n$ are radially arranged, as shown in 
Fig.~\ref{Fig_radial-strips}. 
The $k$th strip carries a transport current of $I_k$, and we consider the 
radial strips carrying unidirectional currents (i.e., all strips carry 
transport currents of identical direction, $I_{k+1}=I_k$) and 
those carrying bidirectional currents (i.e., neighboring strips carry 
transport currents of opposite directions, $I_{k+1}=-I_k$). 
Radial strips of $n=2$ corresponds to two coplanar strips, as investigated 
in Refs.~\onlinecite{Ainbinder03} and \onlinecite{Brojeny02}.


First, we consider radial strips carrying unidirectional currents, 
and each strip carries a transport current $I_k=I_0$ (where $I_k$ is 
monotonically increased from $I_k=0$ to $I_k=I_0$) along the $z$ axis. 
The sheet current in each strip is identical and is denoted by $K_z(\rho)$, 
and the magnetic field component perpendicular to the $k$th strip at 
$\theta=\vartheta_k$ is also identical and is denoted by $H_{\theta}(\rho)$. 
The Biot-Savart law for radially arranged strips with unidirectional currents 
is reduced to 
\begin{equation}
  H_{\theta}(\rho)= \frac{1}{2\pi} \int_b^a d\varrho K_z(\varrho) 
    \frac{n\rho^{n-1}}{\rho^n -\varrho^n} . 
\label{rad-uniI_BS-H}
\end{equation}
In the ideal Meissner state, the perpendicular magnetic field 
in the strips is zero. 
The field distributions (i.e., the sheet current and the perpendicular 
magnetic field) in the ideal Meissner state are given by 
$K_z(\rho)=K_M(\rho;a,b)$ and $H_{\theta}(\rho)=H_M(\rho;a,b)$, 
respectively, and the Meissner distribution functions are given by 
\begin{eqnarray}
  K_M(\rho;u,v) &=& 
  \begin{cases}
    2{\cal H}_0(\rho;u,v) & \mbox{for } v<\rho<u , \\
    0 & \mbox{otherwise,} 
  \end{cases}
\label{rad-I_K-Meissner}\\
  H_M(\rho;u,v) &=& 
  \begin{cases}
    -{\cal H}_0(\rho;u,v) & \mbox{for } 0<\rho<v , \\
    0 & \mbox{for } v<\rho<u , \\
  {\cal H}_0(\rho;u,v) & \mbox{for } \rho>u , 
  \end{cases}
\label{rad-I_H-Meissner}
\end{eqnarray}
where $u>v>0$ and ${\cal H}_0(\rho;u,v)$ is defined by 
\begin{equation}
  {\cal H}_0(\rho;u,v)= \frac{I_0}{2\pi} 
    \frac{n\rho^{n-1}}{\sqrt{|(\rho^n-u^n)(\rho^n-v^n)|}} . 
\label{rad-uniI_H0}
\end{equation}
Equation~\eqref{rad-I_K-Meissner} with \eqref{rad-uniI_H0} corresponds 
to the current distribution in a strip situated in a wedge-shaped 
magnetic cavity.~\cite{Genenko00} 

In the critical state model, $K_z(\rho)$ and $H_{\theta}(\rho)$ satisfy 
\begin{eqnarray}
  K_z(\rho) &=& j_cd 
  \quad\mbox{for } b<\rho<\beta \mbox{ or } \alpha<\rho<a , 
\label{rad_Kz-critical}\\
  H_{\theta}(\rho) &=& 0 
  \hspace{1.8em}\mbox{for } \beta<\rho<\alpha , 
\label{rad_Hy-critical}
\end{eqnarray}
where $\alpha$ and $\beta$ are parameters for the flux front.
We adopt the weight-function approach~\cite{Mikheenko93} to obtain the 
field distributions; 
that is, the $K_z(\rho)$ and $H_{\theta}(\rho)$ in the critical state model 
are given as the superposition of the Meissner distribution functions as 
\begin{eqnarray}
  K_z(\rho) &=& \int_{\alpha}^{a}du\int_{b}^{\beta}dv\, 
    \psi(u,v) K_M(\rho;u,v) , 
\label{rad_Kz-weight}\\
  H_{\theta}(\rho) &=& \int_{\alpha}^{a}du\int_{b}^{\beta}dv\, 
    \psi(u,v) H_M(\rho;u,v) , 
\label{rad_Hy-weight}
\end{eqnarray}
where $\psi(u,v)$ is the weight function which fulfills 
\begin{equation}
  \int_{\alpha}^{a}du\int_{b}^{\beta}dv\,\psi(u,v) =1 . 
\label{rad_weight=1}
\end{equation}
Equations~\eqref{rad_Kz-weight} and \eqref{rad_Hy-weight} satisfy 
Eqs.~\eqref{rad-uniI_BS-H} and \eqref{rad_Hy-critical} for any $\psi(u,v)$, 
and Eq.~\eqref{rad_weight=1} is necessary to guarantee 
$I_0= \int_b^a d\rho K_z(\rho)$. 
We obtain $\alpha$, $\beta$, and $\psi(u,v)$ by solving 
Eqs.~\eqref{rad_Kz-critical}, \eqref{rad_Kz-weight}, 
and \eqref{rad_weight=1}. 

For a small current of $I_0\ll I_c$, the flux fronts are close to the 
edges of the strips, $\alpha\simeq a$ and $\beta\simeq b$, and the weight 
function can be approximated as 
\begin{eqnarray}
  \psi(u,v)\simeq \psi_0/\sqrt{(a-u)(v-b)} , 
\label{rad_psi-approx}
\end{eqnarray}
where $\psi_0$ is a constant. 
Equation~\eqref{rad-I_K-Meissner} with \eqref{rad-uniI_H0} for 
$\alpha\simeq a$ and $\beta\simeq b$ is reduced to 
\begin{equation}
  K_M(\rho;u,v) \simeq 
  \begin{cases}
    2\varphi_b/\sqrt{\rho-v} & \mbox{for } b<v<\rho<\beta , \\
    2\varphi_a/\sqrt{u-\rho} & \mbox{for } \alpha<\rho<u<a , 
  \end{cases}
\label{rad-uniI_K-Meissner_ab}
\end{equation}
where the constants $\varphi_a$ and $\varphi_b$ are defined by 
\begin{equation}
  \varphi_a= \frac{I_0}{2\pi} \sqrt{\frac{na^{n-1}}{a^n-b^n}} , \quad
  \varphi_b= \frac{I_0}{2\pi} \sqrt{\frac{nb^{n-1}}{a^n-b^n}} . 
\label{rad-uniI_cacb}
\end{equation}
Substitution of Eqs.~\eqref{rad_Kz-weight}, \eqref{rad_psi-approx}, 
and \eqref{rad-uniI_K-Meissner_ab} into Eq.~\eqref{rad_Kz-critical} 
yields 
\begin{equation}
  j_cd \simeq 4\pi\psi_0 \varphi_b\sqrt{a-\alpha} , \quad
  j_cd \simeq 4\pi\psi_0 \varphi_a\sqrt{\beta-b} . 
\label{rad-uniI_psi0-ab}
\end{equation}
The constant $\psi_0$ is determined by substituting 
Eqs.~\eqref{rad_psi-approx} and \eqref{rad-uniI_psi0-ab} into 
Eq.~\eqref{rad_weight=1}, yielding $\psi_0=(j_cd)^2/(4\pi^2\varphi_a\varphi_b)$. Equation~\eqref{rad-uniI_psi0-ab} is thus reduced to 
\begin{equation}
  a-\alpha \simeq (\pi \varphi_a/j_cd)^2 , \quad
  \beta-b \simeq (\pi \varphi_b/j_cd)^2 . 
\label{rad-uniI_ab-I0}
\end{equation}
Equation~\eqref{rad-I_H-Meissner} with \eqref{rad-uniI_H0} for 
$\alpha\simeq a$ and $\beta\simeq b$ is reduced to 
\begin{equation}
  H_M(\rho;u,v) \simeq 
  \begin{cases}
    -\varphi_b/\sqrt{v-\rho} & \mbox{for } b<\rho<v<\beta , \\
    \varphi_a/\sqrt{\rho-u} & \mbox{for } \alpha<u<\rho<a , 
  \end{cases}
\label{rad-uniI_H-approx-ab}
\end{equation}
and the perpendicular magnetic field near the edges of strips is calculated 
from Eq.~\eqref{rad_Hy-weight} as 
\begin{equation}
  H_{\theta}(\rho) \simeq 
  \begin{cases}
  \displaystyle -\frac{j_cd}{\pi} \mbox{arctanh}
    \left(\sqrt{\frac{\beta-\rho}{\beta-b}}\right) 
      & \mbox{for } b<\rho<\beta , \\
  \displaystyle \frac{j_cd}{\pi} \mbox{arctanh}
    \left(\sqrt{\frac{\rho-\alpha}{a-\alpha}}\right) 
      & \mbox{for } \alpha<\rho<a . 
  \end{cases}
\label{rad-uniI_Hy-ab}
\end{equation}

Here, we consider radial strips carrying unidirectional ac of 
amplitude $I_0$, and the ac loss in radially arranged strips of 
number $n$ per unit length for one ac cycle is defined by 
$Q =n\int_{\rm cycle}dt\int_{b}^{a}d\rho E_zK_z$, where $E_z$ is the 
electric field in the strips. 
The hysteretic ac loss is calculated from the perpendicular magnetic 
field given by 
Eq.~\eqref{rad-uniI_Hy-ab} for monotonically increased currents as 
\begin{eqnarray}
  Q &=& 4n\mu_0j_cd \left[ \int_{\alpha}^{a} d\rho 
    (a-\rho)H_{\theta}(\rho) \right. 
\nonumber\\
  && \left.\qquad\qquad {}-\int_{b}^{\beta} d\rho 
    (\rho-b)H_{\theta}(\rho) \right] , 
\label{rad_Q-def}
\end{eqnarray}
and substitution of Eqs.~\eqref{rad-uniI_Hy-ab} and \eqref{rad-uniI_ab-I0} 
into Eq.~\eqref{rad_Q-def} yields 
\begin{eqnarray}
  Q &\simeq& \frac{4n}{3\pi}\mu_0(j_cd)^2 
    \Bigl[ (a-\alpha)^2 +(\beta-b)^2 \Bigr] 
\nonumber\\
  &\simeq& \frac{4\pi^3\mu_0 n}{3(j_cd)^2} 
    \Bigl(\varphi_a^{\,4} +\varphi_b^{\,4}\Bigr) . 
\label{rad_Q-ab}
\end{eqnarray}
As seen from Eqs.~\eqref{rad-uniI_K-Meissner_ab}, 
\eqref{rad-uniI_H-approx-ab}, and \eqref{rad_Q-ab}, 
the ac loss for small-current limit ($I_0\ll I_c$) is directly related 
to the field distributions in the ideal Meissner state.~\cite{Kajikawa05} 
The resulting expression for the ac loss in radially arranged strips 
carrying unidirectional currents is obtained by substituting 
Eq.~\eqref{rad-uniI_cacb} into Eq.~\eqref{rad_Q-ab}, yielding 
\begin{equation}
  Q \simeq \frac{\mu_0}{6\pi} \frac{I_0^4}{I_c^2}\, 
    q_n\left(\frac{b}{a}\right) , 
\label{rad-I_Q-res}
\end{equation}
where $\mu_0I_0^4/6\pi I_c^2$ corresponds to the ac loss in a single 
isolated strip~\cite{Norris70} for $I_0\ll I_c$, and the geometrical 
factor $q_n(b/a)$ is defined by 
\begin{equation}
  q_n(s) = \frac{n^3(1-s)^2[1+s^{2(n-1)}]}{2(1-s^n)^2} . 
\label{rad-uniI_qns}
\end{equation}

\begin{figure}[tb] 
\includegraphics{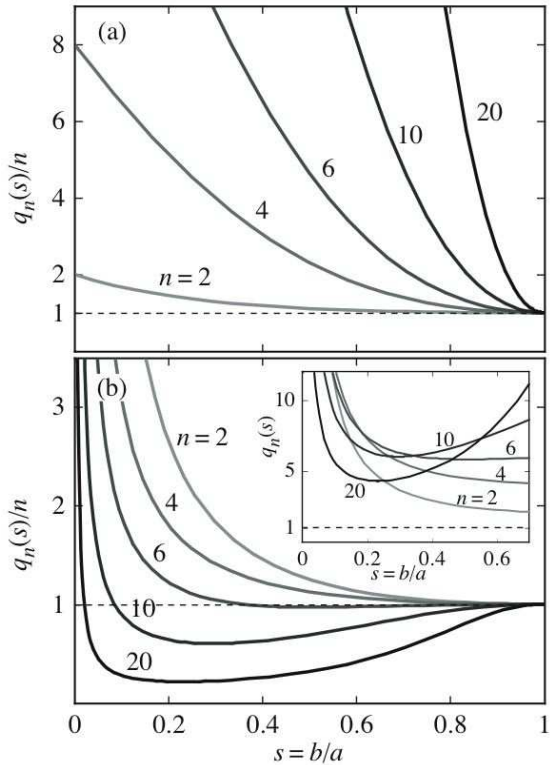}
\caption{%
Geometrical factor $q_n(s)/n$ for ac loss in each strip $Q/n$ as a 
function of $s=b/a$: (a) $q_n(s)/n$ for unidirectional currents 
and (b) $q_n(s)/n$ for bidirectional currents. 
Inset shows the geometrical factor for total loss $q_n(s)$ for 
bidirectional currents. 
}\label{Fig_ac-loss-qns}
\end{figure}

Figure~\ref{Fig_ac-loss-qns} shows the geometrical factor of hysteretic 
ac loss in radial strips as a function of the parameter $s=b/a$ 
and $n$. 
The asymptotic behavior of $q_n(s)/n\to 1$ for $s\to 1$ means that 
electromagnetic interaction between strips can be neglected 
when the spacing between strips becomes much larger than the width 
of the strip. 
The $q_n(s)/n$ for unidirectional currents shown 
in Fig.~\ref{Fig_ac-loss-qns}(a) {\em monotonically decreases} with 
increasing $s$, and $q_n(s)\to n^3/2$ for $s\to 0$ is {\em finite}. 
The $q_n(s)/n$ {\em increases} with $n$, because the magnetic field 
near the outer edges at $\rho\sim a$ increases with $n$. 
For large $n$ (i.e., $s^n\ll 1$), Eq.~\eqref{rad-uniI_qns} 
is reduced to $q_n(s)\simeq n^3(1-s)^2/2\propto n^3$. 
The ratio of the ac loss $Q_b$ arising from the inner edges at 
$\rho\simeq b$ to the ac loss $Q_a$ arising from the outer edges 
at $\rho\simeq a$ is given by $Q_b/Q_a\simeq (\beta-b)^2/(a-\alpha)^2 
\simeq (\varphi_b/\varphi_a)^4= (b/a)^{2(n-1)}$, 
thus yielding $Q_a\geq Q_b$. 

Next, we consider radial strips of {\em even} number $n$ carrying 
bidirectional currents; that is, the $k$th strip carries a transport 
current $I_k=(-1)^kI_0$ along the $z$ axis (where $|I_k|$ is 
monotonically increased from $0$ to $I_0$). 
The sheet current and the perpendicular magnetic field in the $k$th 
strip at $\theta=\vartheta_k$ are given by $(-1)^k K_z(\rho)$ and 
$(-1)^kH_{\theta}(\rho)$, respectively. 
The Biot-Savart law for radially arranged strips with bidirectional currents 
is reduced to 
\begin{equation}
  H_{\theta}(\rho)= \frac{1}{2\pi} \int_b^a d\varrho K_z(\varrho) 
    \frac{n\rho^{n/2-1}\varrho^{n/2}}{\rho^n -\varrho^n} . 
\label{rad-biI_BS-H}
\end{equation}
The field distributions in the ideal Meissner state are given by 
$K_z(\rho)=K_M(\rho;a,b)$ and $H_{\theta}(\rho)=H_M(\rho;a,b)$, 
respectively, and the Meissner distribution functions are given by 
Eqs.~\eqref{rad-I_K-Meissner} and \eqref{rad-I_H-Meissner}, 
where ${\cal H}_0(\rho;u,v)$ for bidirectional currents is defined by 
\begin{equation}
  {\cal H}_0(\rho;u,v)= m_0(u,v) 
    \frac{\rho^{n/2-1}}{\sqrt{\left|(\rho^n-u^n)(\rho^n-v^n)\right|}} , 
\label{rad-biI_H0}
\end{equation}
instead of Eq.~\eqref{rad-uniI_H0} for unidirectional currents. 
The constant $m_0(a,b)$ in Eq.~\eqref{rad-biI_H0} is determined 
such that $I_0=\int_b^a d\rho K_M(\rho;a,b)$, and is given by 
\begin{equation}
  m_0(a,b)= \frac{I_0\, na^{n/2}}{4 \bm{K}(\sqrt{1-b^n/a^n})} , 
\label{rad-biI_m0}
\end{equation}
where $\bm{K}(\kappa)$ is the complete elliptic integral of the first kind. 

The calculations of the field distributions and ac loss for bidirectional 
currents in the critical state model are similar to those for 
unidirectional currents. 
Equations~\eqref{rad_Kz-critical}--\eqref{rad_Q-ab} are all valid 
also for bidirectional currents, except for Eq.~\eqref{rad-uniI_cacb}. 
Instead, the constants $\varphi_a$ and $\varphi_b$ for bidirectional currents 
are defined by 
\begin{equation}
  \varphi_a= \frac{m_0(a,b)}{\sqrt{na(a^n-b^n)}} , \quad
  \varphi_b= \frac{m_0(a,b)}{\sqrt{nb(a^n-b^n)}} . 
\label{rad-biI_cacb}
\end{equation}
The ac loss for bidirectional currents can therefore be obtained by 
substituting Eqs.~\eqref{rad-biI_m0} and \eqref{rad-biI_cacb} into 
Eq.~\eqref{rad_Q-ab}. 
The resulting expression of ac loss in radially arranged strips 
carrying bidirectional currents is given by Eq.~\eqref{rad-I_Q-res}, 
where the geometrical factor $q_n(b/a)$ is given by 
\begin{equation}
  q_n(s) = \frac{\pi^4 n^3(1-s)^2(1+s^2)}{32s^2(1-s^n)^2 
    [\bm{K}(\sqrt{1-s^n})]^4} . 
\label{rad-biI_qns}
\end{equation}

Behavior of the ac loss for bidirectional currents differs significantly 
from that for unidirectional currents. 
The $q_n(s)$ for bidirectional currents shown in 
Fig.~\ref{Fig_ac-loss-qns}(b) is {\em not a monotonic function} of $s$ 
for $n\geq 6$, whereas $q_n(s)$ decreases with increasing $s$ for $n=2$ 
or $4$. 
Because the magnetic field tends to be canceled out, loss per strip 
$q_n(s)/n$ {\em decreases} with $n$, and even the total loss $q_n(s)$ can 
{\em decrease}, as shown in the inset in Fig.~\ref{Fig_ac-loss-qns}(b). 
The elliptic integral in Eq.~\eqref{rad-biI_qns} is reduced to 
$\bm{K}(\sqrt{1-s^n})\simeq \ln(4s^{-n/2})$ for $s^n\ll 1$, 
thus yielding $q_n(s)\propto 1/n$ for $n\to\infty$. 
The $q_n(s)\propto s^{-2}[\ln(s)]^{-4}$ {\em diverges} when $s\to 0$, 
because the magnetic field is concentrated near the inner edges at 
$\rho\sim b$. 
The ratio of the ac loss arising from the inner edges to the ac loss 
arising from the outer edges is given by $Q_b/Q_a\simeq 
(\varphi_b/\varphi_a)^4= (a/b)^2$, thus yielding $Q_a\leq Q_b$.

In summary, we theoretically investigated hysteretic ac loss in 
radially arranged strips based on the critical state model for 
small-current limit, $I_0\ll I_c$. 
The ac loss in radial strips of unit length for one ac cycle is given by 
Eq.~\eqref{rad-I_Q-res} and the geometrical factor $q_n(s)$ is given by 
Eq.~\eqref{rad-uniI_qns} for unidirectional currents or by 
Eq.~\eqref{rad-biI_qns} for bidirectional currents. 
The radial strips carrying bidirectional currents have advantages 
for power application: both leakage magnetic field and ac loss are 
reduced when the number of strips is large. 
Such radial strips might be applied to current leads, fault current 
limiters, and/or cables.\cite{Kajikawa06}

We gratefully acknowledge M. Furuse, S. Fuchino, and H. Yamasaki 
for stimulating discussions. 


%
%

\end{document}